\documentclass{iopart}

\usepackage{amssymb}

\def\be{\begin{equation}}
\def\ee{\end{equation}}
\def\bea{\begin{eqnarray}}
\def\eea{\end{eqnarray}}

\def\p{\partial}
\def\a{\alpha}

\def\k{\kappa}
\def\l{\lambda}

\def\vfi{\varphi}

\def\pt{{$\cal PT$}}

\def\l2{{$L^2(-\pi ,\pi )$ }}

\newcommand{\Sc}{Schr\"odinger }

\begin{document}

 \large

\title {SUSY transformations between digonalizable and
non-diagonalizable Hamiltonians}

\author{
 Boris F Samsonov
}

\address{Department of Physics, Tomsk State
 University, 36 Lenin Avenue, 634050 Tomsk, Russia}

\begin{abstract}
\baselineskip=16pt
\noindent
Recently (see quant-ph/0503040) an explicit example has been given of a
\pt-symmetric {\em non-diagonalizable} Hamiltonian. In this paper
we show that such Hamiltonians appear as supersymmetric (SUSY) partners of
Hermitian (hence diagonalizable) Hamiltonians and  they can be turned back to
diagonalizable forms by appropriate SUSY transformations.
\end{abstract}

\medskip
\medskip



{\bf 1.}
It is well-known that there exist non-Hermitian Hamiltonians
which cannot be reduced to a diagonal form by the change of
the basis (so called {\em non-diagonalizable
Hamiltonians}, see e.g. \cite{Wong}). To illustrate better our ideas we will consider here
only regular Sturm-Liouville problems.
The set of eigenfunctions of a non-diagonalizable Hamiltonian
 is not complete in corresponding Hilbert space \cite{Naimark,Marchenko}.
The {\em characteristic determinant} has multiple roots.
Together with any eigenfunction with a simple eigenvalue coinciding
with a multiple root of the characteristic determinant
there exists a set of {\em associated}
functions \cite{Naimark,Marchenko}. The linear hull of the eigenfunction and corresponding
set of the associated functions forms for the given value of the energy
 the root subspace (see e.g. \cite{Marchenko}).
 Recently an explicit example of an exactly solvable \pt-symmetric
  non-diagonalizable Hamiltonian was given \cite{prepr}.

We have discovered that supersymmetry (SUSY) transformations
 may
convert an Hermitian (hence digonalizable) Hamiltonian to a
non-diagonalizable Hamiltonian,
which in particular can possess the \pt\ symmetry,
 and vice versa. The
possibility which does not appear in the linear algebra.
It is
related with the possibility to ``create" more than one ``bound
state"
at a given non-degenerate value of the energy.
Since the energy level is non-degenerate the other state cannot be an
eigenfunction of the Hamiltonian but it can be an associated
function.
In the opposite process, when we are ``deleting" an
eigenfunction having a non-zero associated function,
the latter is transformed to a ``real eigenfunction".
This looks like it ``emerges from the background" and, therefore
it may be called ``background eigenfunction".

{\bf 2.}
We have found that the possibility described above appears if
{\em second order} SUSY transformations or higher are used.

Let us consider two ordinary second order differential equations
\be\label{1}
(h_0-E)\psi_E(x)=0\qquad  h_0=-\p_x^2+V_0(x)\qquad x\in [a,b]
\ee
\be\label{2}
(h_1-E)\vfi_E(x)=0\qquad  h_1=-\p_x^2+V_1(x)\qquad x\in [a,b]
\ee
with $a$ and $b$ being finite numbers.
We say that the Hamiltonian $h_1$ is related with $h_0$ by a second order
 SUSY  transformation if\\[1em]
1. \vspace{-2em}
\be\label{V1}
V_1=V_0-2 \left[\log W(u_1,u_2)\right]''
\ee
\[
 W(u_1,u_2)=u_1u_2'-u_1'u_2\ne 0 \qquad  \forall x\in (a,b)
\]
2.\vspace{-2em}
\be
(h_0-\a_{1,2})u_{1,2}(x)=0\,.
\ee
It is known (see e.g. \cite{BS}) that in this case $\vfi_E$ is
related with $\psi_E$ as follows:
\be\label{fiE}
\vfi_E=L\psi_E=W(u_1,u_2,\psi_E)/W(u_1,u_2)\qquad
E\ne \a_1,\a_2
\ee
\be\label{fial}
\vfi_{\a_{1,2}}=u_{2,1}/W(u_1,u_2)\,.
\ee
We are using the symbol $W$ to denote Wronskians
and will everywhere suppose that $W(u_1,u_2)\ne 0$
$\forall x\in (a,b)$.
Equation (\ref{fiE}) holds for any $\psi_E$ from the two
dimensional space $\mbox{ker} (h_0-E)$.
The operator $L$ intertwines the Hamiltonians $h_0$ and $h_1$,
$Lh_0=h_1L$.

Let us suppose that $V_0(x)$ is a real-valued and sufficiently
smooth function for $x\in[a,b]$.
Consider two boundary value problems, that we will denote (I) and (II) respectively,
defined by the equations
(\ref{1}) and (\ref{2}) and the boundary conditions
\be\label{bc1}
\psi_E(a)=\psi_E(b)=0
\ee
\be\label{bc2}
\vfi_E(a)=\vfi_E(b)=0\,.
\ee
It is well-known (see e.g. \cite{Levitan}) that the problem (I)
has only discrete, simple and real spectrum of eigenvalues
$E=E_n$, $n=0,1,2,\ldots$.

We
will now formulate conditions for $u_1$ and $u_2$ leading to a
complex-valued $V_1(x)$ given by (\ref{V1})
with a real and simple spectrum
coinciding with the spectrum of $V_0$ except for one level
 and the  Hamiltonian $h_1$ is non-diagonalizable.

It follows from (\ref{fiE}) and (\ref{1}) that
\be\label{vfiEs}
\vfi_E=\frac{1}{W(u_1,u_2)}\,
\left[({W(u_1,u_2)E+\a_2u_1'u_2-\a_1u_1u_2'})\psi_E+
(\a_1-\a_2)u_1u_2\psi_E'\right]\,.
\ee
It is clear from here that if both $u_1$ and $\psi_E$,
$E\ne\a_1,\a_2$,
 satisfy the boundary
conditions (\ref{bc1}) then $\vfi_E$ given by (\ref{vfiEs})
satisfies the boundary conditions (\ref{bc2}).
The only possibility for $u_1$ to satisfy the zero
boundary conditions is to be an eigenfunction of $h_0$,
$u_1=\psi_{E_k}$, so that it is (up to an inessential phase factor)
real and $\a_1=E_k$,
 which we shall suppose to be the case.
This means that the
Hamiltonian $h_1$ has the same spectrum as $h_0$ except maybe for the
values $\a_1$ and $\a_2$ but since $u_1$ is supposed to satisfy
the boundary conditions (\ref{bc1}), the function $\vfi_{\a_2}$
given in (\ref{fial}) is an eigenfunction of $h_1$ and $E=\a_2$ is
the spectral point for $h_1$.
Remembering that we want to keep the real character of the spectrum of $h_1$ we have
to choose $\a_2$ real also.
So, we choose both $\a_1$ and $\a_2$ to be real
and the function $u_1$ is fixed to be real
but we want to get a complex potential difference defined by
equation (\ref{V1}).
This is
possible if $u_2$ is a complex linear combination of two real
linearly independent solutions of equation (\ref{1}).
Let $\a_2(\ne \a_1)$ also coincides with a spectral point $E_l$ of
$h_0$, $\a_2=E_l$, and
$u_2=\psi_{E_l}+i c \psi_{E_l}^{(2)}$, $c\in \Bbb R$ where $\psi_{E_l}$
satisfies the boundary conditions (\ref{bc1}) and $\psi_{E_l}^{(2)}$ is
any real solution of Eq. (\ref{1}) at $E=E_l$
 linearly independent with $\psi_{E_l}$.
We notice that $u_2(x)\ne 0$ $\forall x\in [a,b]$.

We claim that with $u_1$ and $u_2$ being chosen as it is described
above the potential $V_1$ given in (\ref{V1}) has the spectrum
coinciding with the spectrum of the initial $V_0$ except for the
point $E=\a_1=E_k$ which is absent. At the energy $E=\a_2=E_l$ except
for an eigenfunction of $h_1$ there exists an {\em associated
function} (see e.g. \cite{Naimark,Marchenko} and also \cite{prepr})
 which we will also
 call ``background eigenfunction". It satisfies the inhomogeneous
equation
\be\label{backgr1}
(h_1-E_l)\chi_{E_l}=\vfi_{E_l}\qquad
\chi_{E_l}(a)=\chi_{E_l}(b)=0
\ee
and also the homogeneous one with the squared Hamiltonian
\be\label{backgr2}
(h_1-E_l)^2\chi_{E_l}=0\qquad
\chi_{E_l}(a)=\chi_{E_l}(b)=0\qquad \chi_{E_l}\ne \vfi_{E_l}\,.
\ee
We would like to stress that the set $\left\{\vfi_n\right\}$,
$n=0,1,2,\ldots $; $n\ne k$ ($E_k$ is ``deleted")
is not complete in {$L^2(a ,b )$. To have a
complete set one has to add to this set the function
$\chi_{E_l}$  \cite{Naimark,Marchenko}.

As it was already pointed out all spectral points
$E_n$, $n\ne k,l$ of $h_0$ are spectral points of $h_1$ also.
So, to prove our claim it remains to
 analyze only the points $E=\a_1=E_k$ and $E=\a_2=E_l$.

One of the solutions $\vfi_{E_k}^{(1)}=\vfi_{\a_1}$ of the \Sc equation with
$E=\a_1=E_k$ is given by (\ref{fial}) from which it follows that
$\vfi_{E_k}^{(1)}(a)\ne 0$ and $\vfi_{E_k}^{(1)}(b)\ne 0$.
A solution vanishing at one of the
bounds, for instance at $x=a$
\be
\vfi_{E_k}^{(2)}(x)=
\vfi_{E_k}^{(1)}(x)\int_{a}^x\frac{1}{[\vfi_{E_k}^{(1)}(y)]^2}dy
\ee
does not vanish at the other bound. This means that $E=\a_1=E_k$
is not a spectral point of $h_1$.

To get a solution of the \Sc equation at $E=\a_2=E_l$ one can use
 formula (\ref{vfiEs}) with $\psi_E=\psi_{E_l}$ which gives us the
 function $\vfi_{E_l}$ satisfying the zero boundary conditions
 meaning
as it was already mentionned
 that $E=E_l$ is the spectral point for $h_1$. Moreover,
 since $u_1(a)=0$, from (\ref{vfiEs}) one gets
  $\vfi_E(a)= (E-\a_2)\psi_E(a)$. Now
  if $\psi_E(a)\to\psi_{E_l}(a)$ when $E\to E_l$, remembering that
  $\psi_E(a)$ is an analytic function of $E$ having a simple
  zero at $E=E_l$
(see e.g. \cite{Levitan})
   we conclude that the function $\vfi_E(a)$ is also an analytic
  function of $E$ but it
  has a double zero at $E=\a_2=E_l$. In such a case together with the
  function $\vfi_{E_l}$ there exists an associated function
  $\chi_{E_l}=(\partial\vfi_{E} /\partial_{E})_{E=E_l}$
  (see e.g. \cite{Naimark,Marchenko}).
  It is evident that $\chi_{E_l}(a)=\chi_{E_l}(b)=0$
  and the equation (\ref{backgr1}) it satisfies can be obtained by
   taking the derivative of equation (\ref{2}) with respect to $E$.
   Since $\vfi_E=L\psi_E$ and $L$ is independent of $E$ one has
   $\chi_{E_l}=L\tilde\psi_{E_l}$,
    $\tilde\psi_{E_l}=(\partial\psi_{E} /\partial_{E})_{E=E_l}$.
   The function $\tilde\psi_{E_l}$
   satisfies the equation
$
   (h_0-E_l)\tilde\psi_{E_l}=\psi_{E_l}
$
   but it does not satisfy the zero boundary conditions
   which agrees with the fact that $h_0$ is a diagonalizable
   Hamiltonian.
   Operator $L$ (\ref{fial}) turns $\tilde\psi_{E_l}$
    into a solution of the
    equation (\ref{backgr1}) satisfying the zero
   boundary conditions thus transforming it into a ``background
   eigenfunction" of $h_1$.

In
contrast to the usual SUSY scheme the opposite process,
the ``deletion" of the level $E=E_l$  does not actually delete this
level.
If we take the Hamiltonian $h_1$ as the initial Hamiltonian
for the next second order transformation,
leading to the Hamiltonian $h_2$,
 and choose one of
the transformation functions
defining the transformation operator $L^{(2)}$ of the next step
to be equal to $\vfi_{E_l}$, actual
eigenfunction at $E=E_l$ is deleted but the associated function
$\chi_{E_l}$ ``comes out of the background" and becomes a
 true eigenfunction of $h_2$
  at $E=E_l$.
This statement is readily seen if one acts by $L^{(2)}$,
which is constructed in a similar way as $L=L^{(1)}$ given in
(\ref{fiE})
and intertwines now $h_1$ and $h_2$, on
both sides of Eq.
(\ref{backgr1}), takes into account the intertwining relation
$L^{(2)}h_1=h_2L^{(2)}$
and the property
$L^{(2)}\vfi_{E_l}=0$.
If a non-diagonalizable Hamiltonian  has only one associated function
it is transformed in this way into a diagonalizable Hamiltonian.

{\bf 3.}
The simplest example illustrating the possibilities described
above is the boundary value problem with the zero initial potential
$V_0(x)=0$. We will choose $a=-\pi$ and $b=\pi$. The solutions of
the boundary value problem (\ref{1}), (\ref{bc1}) is well-known, for
instance, its discrete spectrum is $E=E_n=\frac 14n^2$,
$n=1,2,\ldots$.

Let us choose $u_1=\sin(Ax)$ and $u_2=\exp(-iBx)$, $A,B\in \Bbb R$.
Formula (\ref{V1}) gives us the following \pt-symmetric
Hamiltonian:
\be\label{V1ex}
V_1=\frac{2 A^2(A^2-B^2)}{[\,\cos(Ax)-iB\sin(Ax)]^2}\,.
\ee
For $A=1$ the function $u_1$ coincides with the first excited state of
$h_0$ and for $B\ne n/2$ this potential is diagonalizable with the
spectrum $E=E_n=\frac 14n^2$, $n=1,3,4,5,\ldots$ and
$E_{\a_2}=B^2$.
For $B=2$ the function $u_2$ is a complex linear
combination of the fourth excited state and another solution of Eq.
(\ref{1}) with $V_0(x)=0$ at
the same energy and the level $E_{\a_2}$ merges with the existing
level $E=4$ which ``goes to the background".
The potential (\ref{V1ex})
becomes non-diagonalizable
with the discrete spectrum $E=E_n=\frac 14n^2$,
$n=1,3,4,5,\ldots$ studied in detail in
\cite{prepr}.

Now we would like to illustrate the possibility to transform the
non-diagonalizable potential (\ref{V1ex}) at $A=1$ and $B=2$
 into a digonalizable one.
 We choose $V_1$ as the initial potential and take
$u_1=\vfi_4$ and $u_2=\vfi_{left}$
where $\vfi_{left}$ is such that $\vfi_{left}(-\pi)=0$.
This yields us the following potential:
\be\label{V2}
V_2=\frac{(\k^2-1)[\k^2-1-\k^2\cos(2x)+\cos(2\k x+2\k\pi )]}%
{[\k\cos(\k x+\k\pi )\sin x-\sin(\k x+\k\pi)\cos x]^2}
\qquad \k\ne  1
\ee
where we denoted $\a_2=\k^2$.
It is regular $\forall x\in (-\pi ,\pi )$ provided
$0.5\le \k\le 1.5$, $\k\ne 1$
and has the spectrum $E=E_n=\frac{n^2}{4}$, $n=1,3,4\ldots $ and
$E=\a_2=\k^2$.
For $\k= 1$
$\vfi_{left}=L\psi_{left}=0$ and to realize the transformation with
$\a_2=1$ one has to use the solution obtained with the help of
formula (\ref{fial}). This corresponds to the backward
transformation from $V_1$ to $V_0=0$ and hence one gets  $V_2=0$.
For a real $\k$ the potential (\ref{V2}) is real and corresponds to the
Hermitian (hence diagonalizable) Hamiltonian $h_2=-\p_x^2+V_2$.
 So, we have transformed the non-diagonalizable Hamiltonian
$h_1$ to the diagonalizable $h_2$. One can also transform $h_1$
into a non-Hermitian diagonalizable $h_2$ by choosing a complex linear
combination of two linearly independent solutions of equation
(\ref{2}) corresponding to the same value of $E=\a_2$ as transformation
function $u_2$.

Our last example is related with the possibility to enlarge the root
subspace corresponding to $E=4$ of the potential (\ref{V1ex}) at
$A=1$ and $B=2$ from the dimension two till the dimension three.
For this aim we take $u_1=\vfi_{E_1}$ and
$u_2=[9-\exp(-2ix)]/[1-3\exp(2ix)]$ which
yields the potential
\[
V_2=6\,\frac{25e^{ix}+324e^{2ix}+1350e^{3ix}+2500e^{4ix}+2025e^{5ix}}%
{(3+25e^{ix}+81e^{2ix}+75e^{3ix})^2}\,.
\]
It has the spectrum $E=\frac{n^2}{4}$, $n=3,4,5,\ldots$.

We hope that the possibility to transform non-diagonalizable
\pt-symmetric Hamiltonians to diagonalizable forms may find
application in  complex quantum mechanics which is currently under
developments.

This work is partially supported by the
President Grant of Russia 1743.2003.2
 and the Spanish MCYT and European FEDER grant BFM2002-03773.
The author would like to acknowledge the hospitality of Physics and
Applied Mathematics Unit of the Indian Statistical Institute
 in winter 2004 where this work has been started.

\end{document}